\documentclass[twocolumn,tighten,twocolappendix]{aastex63}
\pdfoutput=1 
\usepackage{amsmath,amstext}
\usepackage[T1]{fontenc}
\usepackage{ae,aecompl}
\usepackage[utf8]{inputenc}
\usepackage{apjfonts} 
\usepackage[figure,figure*]{hypcap}
\usepackage{enumitem}
\usepackage{bm}
\usepackage{pifont}
\usepackage{bigints}
\usepackage{lineno}
\input{hyperlink-year-only-natbib-patch}

\makeatletter
\DeclareRobustCommand{\H}{%
  \mbox{H\check@mathfonts\fontsize\sf@size\z@\selectfont I}%
}
\makeatother

\newcommand{\HH}{${\rm H_2}$}

\graphicspath{{./}{figures/}}

\shorttitle{\HH\ Galaxy Formation Threshold}
\shortauthors{E.~O.~Nadler}

\begin{document}

\title{The Impact of Molecular Hydrogen Cooling on the Galaxy Formation Threshold}

\correspondingauthor{Ethan~O.~Nadler}
\email{enadler@ucsd.edu}
\author[0000-0002-1182-3825]{Ethan~O.~Nadler}
\affiliation{Department of Astronomy \& Astrophysics, University of California, San Diego, La Jolla, CA 92093, USA}

\begin{abstract}
We study the impact of molecular (\HH) and atomic (\H) hydrogen cooling on the galaxy formation threshold. We calculate the fraction of dark matter (DM) halos that exceeds a critical mass required for star formation, $M_{\mathrm{crit}}(z)$, as a function of their peak mass. By convolving analytic halo mass accretion histories (MAHs) with models for $M_{\mathrm{crit}}(z)$, we predict that halos with peak virial masses below $\sim 10^8~M_{\mathrm{\odot}}$ can form stars before reionization through \HH\ cooling. These halos remain dark when only \H\ cooling and reionization are modeled. However, less than $\approx 10\%$ of halos with peak masses below $\sim 10^{7}~M_{\mathrm{\odot}}$ ever exceed $M_{\mathrm{crit}}(z)$, even when \HH\ cooling is included; this threshold is primarily set by relative streaming motion between DM and baryons imprinted at recombination. We obtain similar results using subhalo MAHs from an extremely high-resolution cosmological DM--only zoom-in simulation of a Milky Way (MW) analog (particle mass $6.3\times 10^3~M_{\mathrm{\odot}}$). Based on the abundance of MW satellites, these results imply that at least some known ultrafaint dwarf galaxies formed through \HH\ cooling. This work sharpens predictions for the galaxy formation threshold and demonstrates how its essential features emerge from the underlying distribution of halo growth histories.
\end{abstract}

\keywords{\href{http://astrothesaurus.org/uat/595}{Galaxy formation (595)}; 
\href{http://astrothesaurus.org/uat/1569}{Star formation (1569)};
\href{http://astrothesaurus.org/uat/1880}{Galaxy dark matter halos (1880)}
}

\section{Introduction}
\label{sec:intro}

What is the halo mass threshold for galaxy formation? This question underlies key areas of research in galaxy formation and cosmology, including when and how the first galaxies formed, which galaxies drive cosmic reionization, and whether ``dark'' (galaxy-free) dark matter (DM) halos exist. Robust predictions for the galaxy formation threshold are critical given upcoming observations of faint galaxies and low-mass halos throughout cosmic history (e.g., see \citealt{Bechtol220307354} and \citealt{Robertson211013160} for reviews).

The physics that regulates galaxy formation in low-mass halos is well studied. Molecular hydrogen (\HH) cooling enables star formation in $\sim 10^6~M_{\mathrm{\odot}}$ `minihalos' before reionization ($z\gtrsim 10$; \citealt{Haiman9511126,Tegmark9603007}). Hydrodynamic simulations indicate that \HH\ cooling is partially---but not entirely---suppressed by Lyman--Werner (LW) background radiation at these redshifts (e.g., \citealt{Machacek0007198,Abel0112088,Wise07072059}). Meanwhile, atomic hydrogen (\H) is the main coolant in halos with masses above $\sim 10^8~M_{\mathrm{\odot}}$, or virial temperatures above $\sim 10^4~\mathrm{K}$, before reionization \citep{Greif08032237}. Reionization subsequently heats the intergalactic medium (IGM), raising the minimum halo mass for star formation above $\sim 10^9~M_{\mathrm{\odot}}$ at $z\lesssim 5$ \citep{Efstathiou1992,Barkana9901114}.

Translating these processes into a prediction for which halos host galaxies across cosmic time remains challenging. Before reionization, halos more massive than the \H\ cooling limit are expected to form stars \citep{Greif08032237}. However, \HH\ cooling can occur in halos above and below this limit, with an efficiency set by the local, time-dependent LW background (e.g., \citealt{Haiman9608130,Oh0108071,OShea07064416}). The first galaxies drive reionization, which sets the star formation threshold at later times (e.g., \citealt{Wise14036123,Norman170500026}). Cosmological hydrodynamic simulations are needed to capture how Population III star formation, external enrichment and metal-line cooling, self-shielding, cold-mode accretion, supernova feedback, and DM--baryon streaming affect this picture (e.g., \citealt{Wise10112632,Jaacks180407372,Schauer181112920,Schauer200005663,Liu200615260,Skinner200104480,Hicks200905499,Hicks240720429,Kulkarni201004169,Munshi210105822}).

Despite this complexity, several recent studies predict a global, time-dependent halo mass threshold for star formation, $M_{\mathrm{crit}}(z)$, by combining analytic calculations and semianalytic models fit to hydrodynamic simulations. Specifically, \defcitealias{Nebrin230308024}{NGM23} Nebrin et al.\ (\citeyear{Nebrin230308024}, hereafter \citetalias{Nebrin230308024}) modeled the impact of \HH\ and \H\ cooling on the galaxy formation threshold, accounting for LW radiation, self-shielding, reionization, and DM--baryon streaming; also see \defcitealias{Hegde230403358}{HF23} Hegde \& Furlanetto (\citeyear{Hegde230403358}, hereafter \citetalias{Hegde230403358}). Meanwhile, \defcitealias{Benitez-Llambay200406124}{BF20} Benitez-Llambay \& Frenk (\citeyear{Benitez-Llambay200406124}, hereafter \citetalias{Benitez-Llambay200406124}) focused on the effects of \H\ cooling and reionization without modeling \HH\ cooling. \citetalias{Benitez-Llambay200406124} convolved their model for $M_{\mathrm{crit}}(z)$ with halo mass accretion histories (MAHs) from extended Press--Schechter (ePS) theory and cosmological simulations to predict the galaxy occupation fraction, $f_{\mathrm{gal}}$: the fraction of DM halos that host galaxies as a function of halo mass.

Herein, we model the impact of \HH\ and \H\ cooling on the galaxy formation threshold, focusing on the lowest-mass halos that form stars before reionization. Unlike previous work, we predict $f_{\mathrm{gal}}$ using models for $M_{\mathrm{crit}}(z)$ with and without \HH\ cooling and DM--baryon streaming. We apply our model to both isolated halos, using analytic MAHs, and subhalos, using MAHs from a high-resolution cosmological zoom-in simulation of a Milky Way (MW) analog. Throughout, we demonstrate how $f_{\mathrm{gal}}$ arises from the underlying distribution of halo growth histories.

These predictions are timely in light of upcoming data. The Vera C.\ Rubin Observatory \citep{IvezicLSST} will discover hundreds of new dwarf galaxies, improving constraints on the faint-end galaxy--halo connection \citep{Nadler240110318}. Predictions for $f_{\mathrm{gal}}$ in various galaxy formation scenarios will be needed to accurately interpret these observations. In parallel, strong gravitational lensing is beginning to probe (sub)halos that may remain dark (see \citealt{Vegetti230611781} for a review). A detection of ``dark'' halos using strong lensing will thus require a thorough understanding of the galaxy formation threshold itself.

This Letter is organized as follows. In Section~\ref{sec:methods}, we describe our $M_{\mathrm{crit}}(z)$ and isolated halo MAH models and our procedure for calculating $f_{\mathrm{gal}}$. We predict $f_{\mathrm{gal}}$ for isolated halos in Section~\ref{sec:results} and for MW subhalos in Section~\ref{sec:sim}. We summarize and discuss our results in Section~\ref{sec:discussion}. We define halo masses using the \cite{Bryan9710107} virial overdensity and adopt cosmological parameters  $h = 0.7$, $\Omega_{\rm m} = 0.286$, $\Omega_{\rm b} = 0.047$, and $\Omega_{\Lambda} = 0.714$ \citep{Hinshaw_2013}.

\section{Model}
\label{sec:methods}

\subsection{Critical Halo Mass for Star Formation}

\begin{figure}[t!]
\hspace{-2.75mm}
\includegraphics[trim={0cm 0.3cm 0 0cm},width=0.48\textwidth]{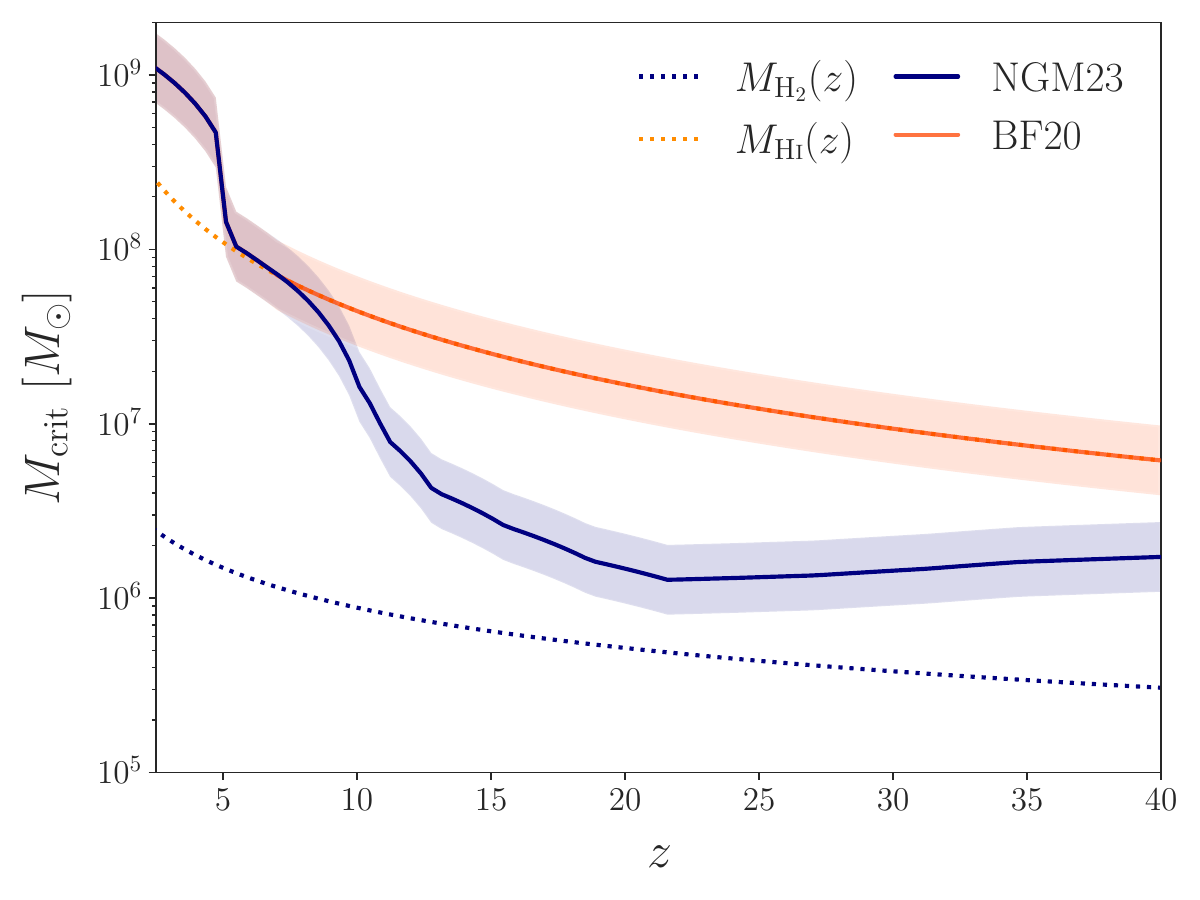}
\caption{Halo mass thresholds for star formation. The thick blue line (which includes \HH\ cooling) shows our fiducial \citetalias{Nebrin230308024} model with DM--baryon streaming velocity $v_{\mathrm{str}}(z)=\sigma_{\mathrm{str}}(z)$. The thick orange line (which does not include \HH\ cooling) shows our fiducial \citetalias{Benitez-Llambay200406124} model with $z_{\mathrm{reion}}=7$. Shaded bands show the assumed $0.2~\mathrm{dex}$ scatter in $\log(M_{\mathrm{crit}}(z))$. \HH\ and \H\ cooling-only limits from \citetalias{Nebrin230308024} are shown by dotted blue and orange lines, respectively.}
\label{fig:Mcrit}
\end{figure}

We use $M_{\mathrm{crit}}(z)$ from \citetalias{Nebrin230308024} and \citetalias{Benitez-Llambay200406124} as our fiducial star formation thresholds with and without \HH\ cooling, respectively. We apply a symmetric, redshift-independent scatter of $\sigma_{\log(M)}=0.2~\mathrm{dex}$ to each of these global $M_{\mathrm{crit}}(z)$ histories. This value is comparable to the intrinsic scatter in $M_{\mathrm{crit}}(z)$ measured from small-volume ($\lesssim 1~\mathrm{Mpc}~h^{-1}$) hydrodynamic simulations using a uniform LW background \citep{Kulkarni201004169}. Figure~\ref{fig:Mcrit} shows both $M_{\mathrm{crit}}(z)$ models (solid blue and orange lines and bands), along with the \HH\ and \H\ cooling-only limits from \citetalias{Nebrin230308024} (dotted blue and orange lines).

We implement the \citetalias{Nebrin230308024} model as follows. We choose a fiducial DM--baryon streaming velocity of $v_{\mathrm{str}}(z)=\sigma_{\mathrm{str}}(z)$, where $\sigma_{\mathrm{str}}(z)$ is the global rms value \citep{Tseliakhovich10052416,Fialkov11102111}. We make this choice because the streaming velocity follows a Maxwellian distribution with a most probable
value of $0.82\sigma_{\mathrm{str}}(z)$ (see \citetalias{Hegde230403358}). Based on the properties of MW satellites and metal-poor stars in the MW stellar halo, \cite{Uysal221112838} inferred $v_{\mathrm{str}}(z)\gtrsim \sigma_{\mathrm{str}}(z)$ locally; we study the effects of varying $v_{\mathrm{str}}(z)$ in Appendix~\ref{sec:streaming}. Following \citetalias{Nebrin230308024}, we use the photoionization history from \cite{Faucher190308657} and the LW background from \cite{Incatasciato230108242}, such that \H\ is reionized by $z\approx 7$. Note that LW radiation does not instantly heat self-shielded gas in halos at high redshifts; thus, $M_{\mathrm{crit}}(z)$ increases rapidly at $z\approx 5$ in our \citetalias{Nebrin230308024} implementation. The \citetalias{Nebrin230308024} model we adopt does not include metal-line cooling or dust-catalyzed \HH\ formation, which these authors conclude negligibly affect $M_{\mathrm{crit}}(z)$.

\citetalias{Benitez-Llambay200406124} assumed that \H\ cooling sets $M_{\mathrm{crit}}(z)$ before reionization and that the IGM is then instantaneously heated to $2\times 10^4~\mathrm{K}$, such that $M_{\mathrm{crit}}(z)$ is set by the corresponding virial temperature threshold afterward. Thus, a relevant free parameter in \citetalias{Benitez-Llambay200406124} is the reionization redshift, $z_{\mathrm{reion}}$. We set $z_{\mathrm{reion}}=7$ and match $M_{\mathrm{crit}}(z)$ to \citetalias{Nebrin230308024} for $z<z_{\mathrm{reion}}$. Figure~\ref{fig:Mcrit} shows that our \citetalias{Benitez-Llambay200406124} $M_{\mathrm{crit}}(z)$ model matches the \H\ cooling limit before reionization and matches our \citetalias{Nebrin230308024} model after. We refer the reader to \citetalias{Benitez-Llambay200406124} for a study of how varying $z_{\mathrm{reion}}$ affects $f_{\mathrm{gal}}$ predictions in this model.

\begin{figure*}[t!]
\hspace{-2.25mm}
\includegraphics[trim={0cm 0.3cm 0 0},width=0.5\textwidth]{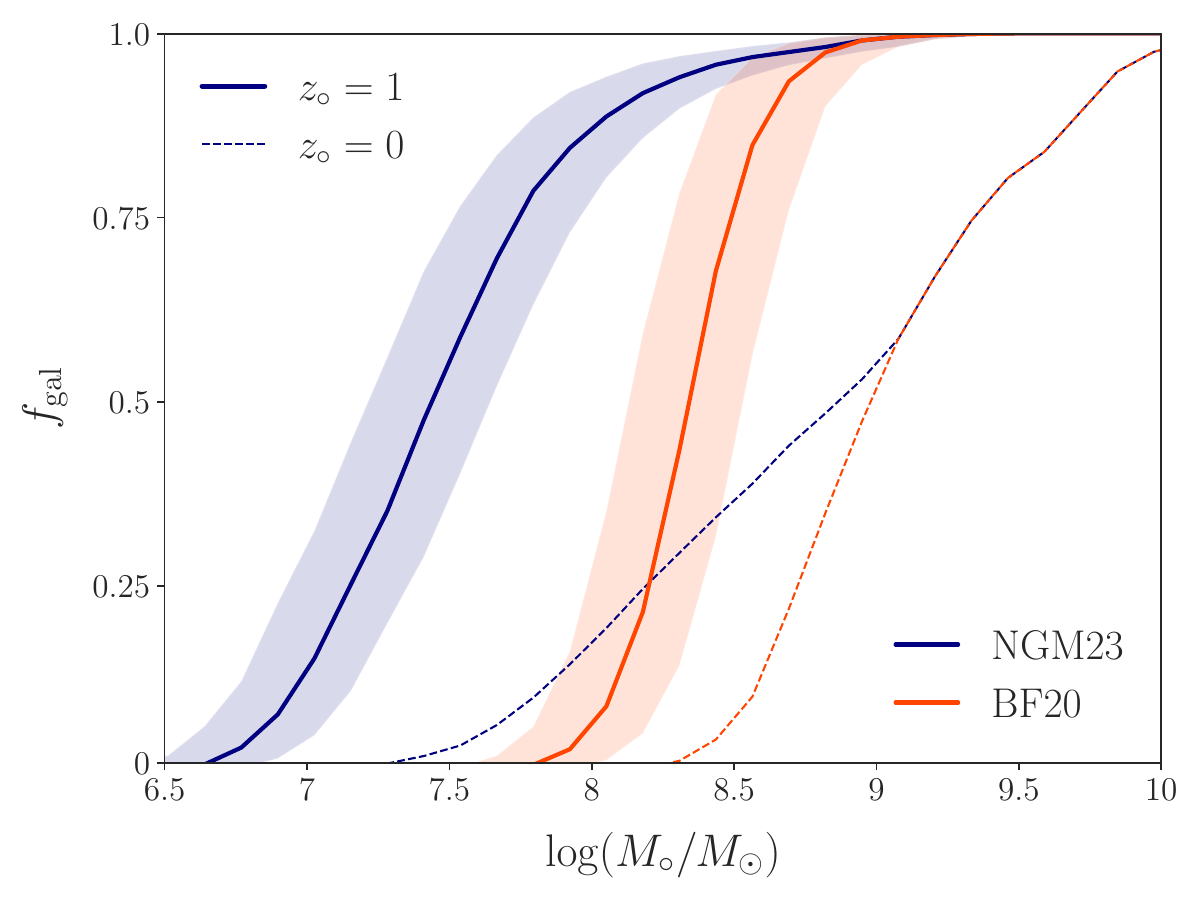}
\hspace{1.25mm}
\includegraphics[trim={0cm 0.3cm 0 0},width=0.5\textwidth]{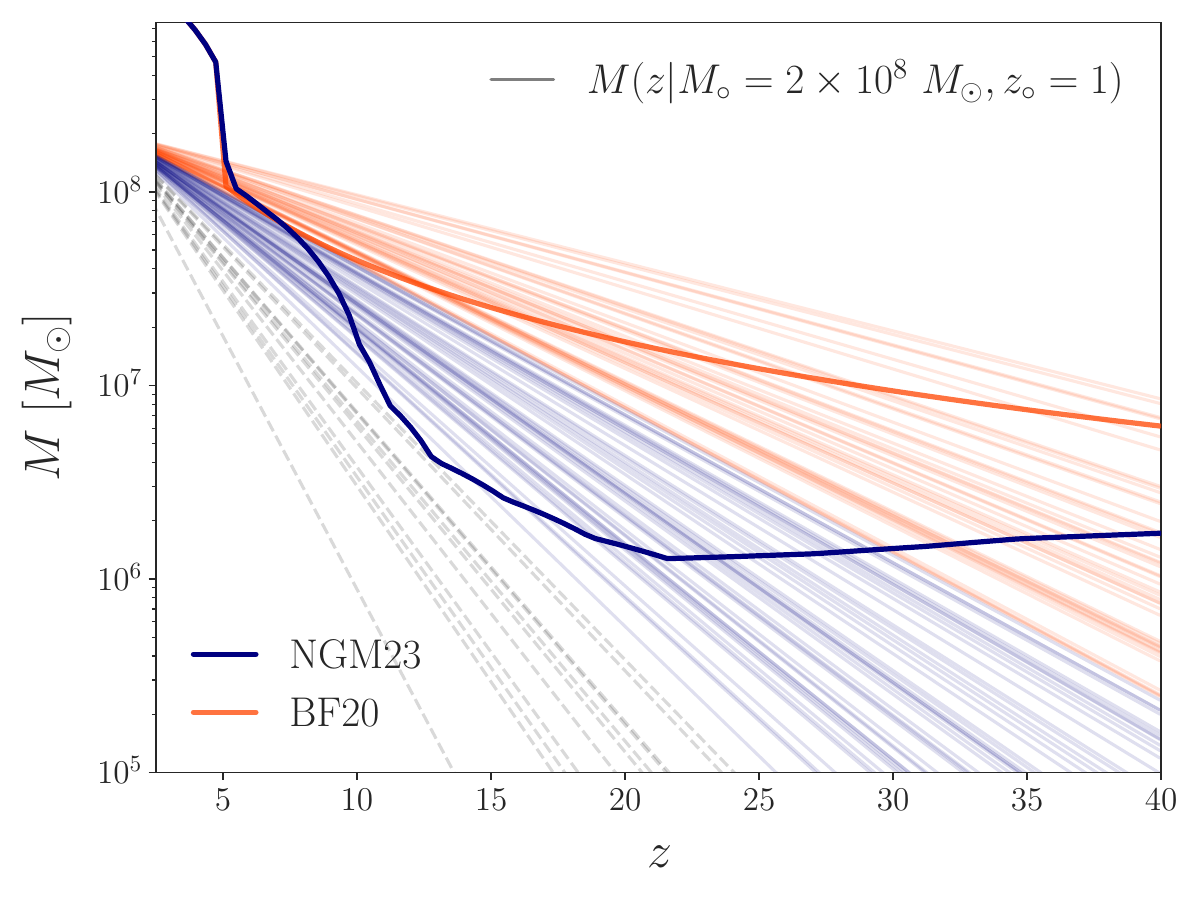}
\caption{\emph{Left}: Median galaxy occupation fraction for isolated halos as a function of halo mass at $z_{\circ}=1$ (solid) and $z_{\circ}=0$ (dashed), generated by combining analytic MAHs with $M_{\mathrm{crit}}(z)$ from \citetalias{Nebrin230308024} (blue) or \citetalias{Benitez-Llambay200406124} (orange). Shaded bands show $16\%$ to $84\%$ intrinsic scatter quantiles. \emph{Right}: Halo growth histories. Thin lines show MAHs with $z_{\circ}=1$ and $M_{\circ}=2\times 10^8~M_{\mathrm{\odot}}$, colored according to whether they exceed the mean $M_{\mathrm{crit}}(z)$ from \citetalias{Benitez-Llambay200406124} (orange), \citetalias{Nebrin230308024} (blue), or neither (dashed gray). Thick lines show the $M_{\mathrm{crit}}(z)$ models.}
\label{fig:fgal_Mcrit}
\end{figure*}

\subsection{Halo Mass Accretion Histories}

We model isolated halo MAHs via \citep{Wechsler0108151}
\begin{align}
    M(z\lvert M_{\circ},z_{\circ}) &= M_{\circ}\exp\left[-\frac{Sc_1}{c_{\circ}(1+z_{\circ})}\left(\frac{1+z}{1+z_{\circ}}-1\right)\right]&\nonumber \\ &=M_{\circ}\exp\left[-\frac{Sc_1}{c_{\circ}}\frac{z-z_{\circ}}{(1+z_{\circ})^2}\right],&\label{eq:M_z}
\end{align}
where $M_{\circ}\equiv M(z_{\circ})$ is the halo mass at redshift $z_{\circ}$, $S=2$, $c_{1}=4.2$ is the virial concentration of a typical halo collapsing today, and $c_{\circ}$ is the concentration at $z_{\circ}$. While Equation~\ref{eq:M_z} was not explicitly calibrated to simulations over the mass and redshift range we study, it provides simple and powerful predictions that are sufficient to capture key features of the galaxy formation threshold. This model has been refined in subsequent work (e.g., \citealt{vandenBosch14092750}), and we leave a detailed exploration of such MAH models to future work.

To generate MAHs, we sample $c_{\circ}$ from the \cite{Diemer180907326} mass--concentration relation at $z_{\circ}$ with a redshift-independent scatter of $\sigma_{\log(c)}=0.16~\mathrm{dex}$. We study the effects of varying $\sigma_{\log(c)}$ in Appendix~\ref{sec:sigma_logc}.

\subsection{Galaxy Occupation Fraction}

We define the galaxy occupation as follows:
\begin{align}
    &f_{\mathrm{gal}}(M_{\circ},z_{\circ}) \equiv \int \mathrm{d}\Delta \,\big[ p(\Delta) \times& \nonumber \\
    &p\left(\exists z \in [z_{\mathrm{min}},z_{\mathrm{max}}]: 
    \log(M(z|M_{\circ},z_{\circ}))
    > \log(M_{\mathrm{crit}}(z)) + \Delta\right)\big],&\label{eq:fgal}
\end{align}
where we marginalize over scatter in $\log(M_{\mathrm{crit}}(z))$ by drawing from $p(\Delta)$ as described in each section below. We set $z_{\mathrm{max}}=40$ and do not model star formation at earlier times. Note that DM--baryon streaming causes $M_{\mathrm{crit}}(z)$ to increase for $z\gtrsim 20$ in our fiducial \citetalias{Nebrin230308024} model; as a result, our predictions in this case are not sensitive to $z_{\mathrm{max}}$, so long as $z_{\mathrm{max}}\gtrsim 20$. We set $z_{\mathrm{min}}=z_{\circ}$; increasing $z_{\mathrm{min}}$ does not significantly affect our predictions, so long as $z_{\mathrm{min}}<z_{\mathrm{reion}}$, because $M_{\mathrm{crit}}(z)$ increases sharply at reionization in both $M_{\mathrm{crit}}(z)$ models we consider.

According to Equation~\ref{eq:fgal}, the fraction of halos that host galaxies is equal to the probability that their MAHs exceed $M_{\mathrm{crit}}(z)$ between $z_{\circ}$ and $z_{\mathrm{max}}>z_{\circ}$. Thus, we effectively assume that halos which exceed $M_{\mathrm{crit}}(z)$ for any amount of time form and retain stars. This assumption is simplistic, since star formation is not instantaneous or permanent. Nonetheless, we will show that our predictions broadly agree with more detailed semianalytic models, suggesting that Equation~\ref{eq:fgal} captures the key physics that sets $f_{\mathrm{gal}}$.

\section{Predictions for Isolated Halos}
\label{sec:results}

To calculate $f_{\mathrm{gal}}$ for isolated halos, we generate MAHs in bins of $\log(M_{\circ}/M_{\mathrm{\odot}})$ and marginalize over $\Delta$ in each bin by drawing from a normal distribution $\mathcal{N}(\mu=0,\sigma=\sigma_{\log(M)})$. Thus, we predict both the mean $f_{\mathrm{gal}}$ and its intrinsic scatter. We do not model cosmic variance due to large-scale fluctuations in the photoionization history or LW background.

The left panel of Figure~\ref{fig:fgal_Mcrit} shows our $f_{\mathrm{gal}}$ predictions for isolated halos with $z_{\circ}=0$ (dashed) and $z_{\circ}=1$ (solid) using the \citetalias{Nebrin230308024} (blue) and \citetalias{Benitez-Llambay200406124} (orange) $M_{\mathrm{crit}}(z)$ models. We predict that halos with $M(z=0)\gtrsim 10^{8.5}~M_{\mathrm{\odot}}$ form stars through \H\ cooling, in agreement with \citetalias{Benitez-Llambay200406124} (see their Figure~11). Meanwhile, halos with $10^{7.5}~M_{\mathrm{\odot}}\lesssim M(z=0)\lesssim 10^{8.5}~M_{\mathrm{\odot}}$ can form stars through \HH\ cooling. For $z_{\circ}=1$, \H\ cooling enables star formation for $M(z=1)\gtrsim 10^{8}~M_{\mathrm{\odot}}$ halos, while \HH\ cooling enables star formation for $10^7~M_{\mathrm{\odot}}\lesssim  M(z=1)\lesssim 10^{8}~M_{\mathrm{\odot}}$.\footnote{Note that $z_{\circ}=1$ corresponds to the redshift at which typical MW subhalos reach their peak mass \citep{Kazuno241022185}.} The resulting lower limits are primarily set by DM--baryon streaming, which causes the \citetalias{Nebrin230308024} $M_{\mathrm{crit}}(z)$ model shown in Figure~\ref{fig:Mcrit} to flatten at high redshifts.

At fixed $M_{\circ}$, increasing $z_{\circ}$ raises $f_{\mathrm{gal}}$ because it forces halos to achieve a given mass earlier, resulting in even larger masses at later times. When varying $z_{\circ}$, our $f_{\mathrm{gal}}$ predictions are connected by the underlying distribution of halo MAHs, which is not linear in $z_{\circ}$. As a result, changing $z_{\circ}$ alters the slope of $f_{\mathrm{gal}}$ in addition to shifting it horizontally. In the $z_{\circ}=0$ case, MAHs along the $f_{\mathrm{gal}}$ cutoff intersect the rapidly-rising part of our $M_{\mathrm{crit}}(z)$ models near reionization, yielding a sharper feature in $f_{\mathrm{gal}}$; meanwhile, in the $z_{\circ}=1$ case, MAHs cross $M_{\mathrm{crit}}(z)$ in a regime where it is roughly constant in the \citetalias{Nebrin230308024} case. In Appendix~\ref{sec:derivation}, we derive the error function-like shape of $f_{\mathrm{gal}}$ analytically under certain assumptions.

\begin{figure*}[t!]
\hspace{-2.25mm}
\includegraphics[trim={0cm 0.3cm 0 0},width=0.5\textwidth]{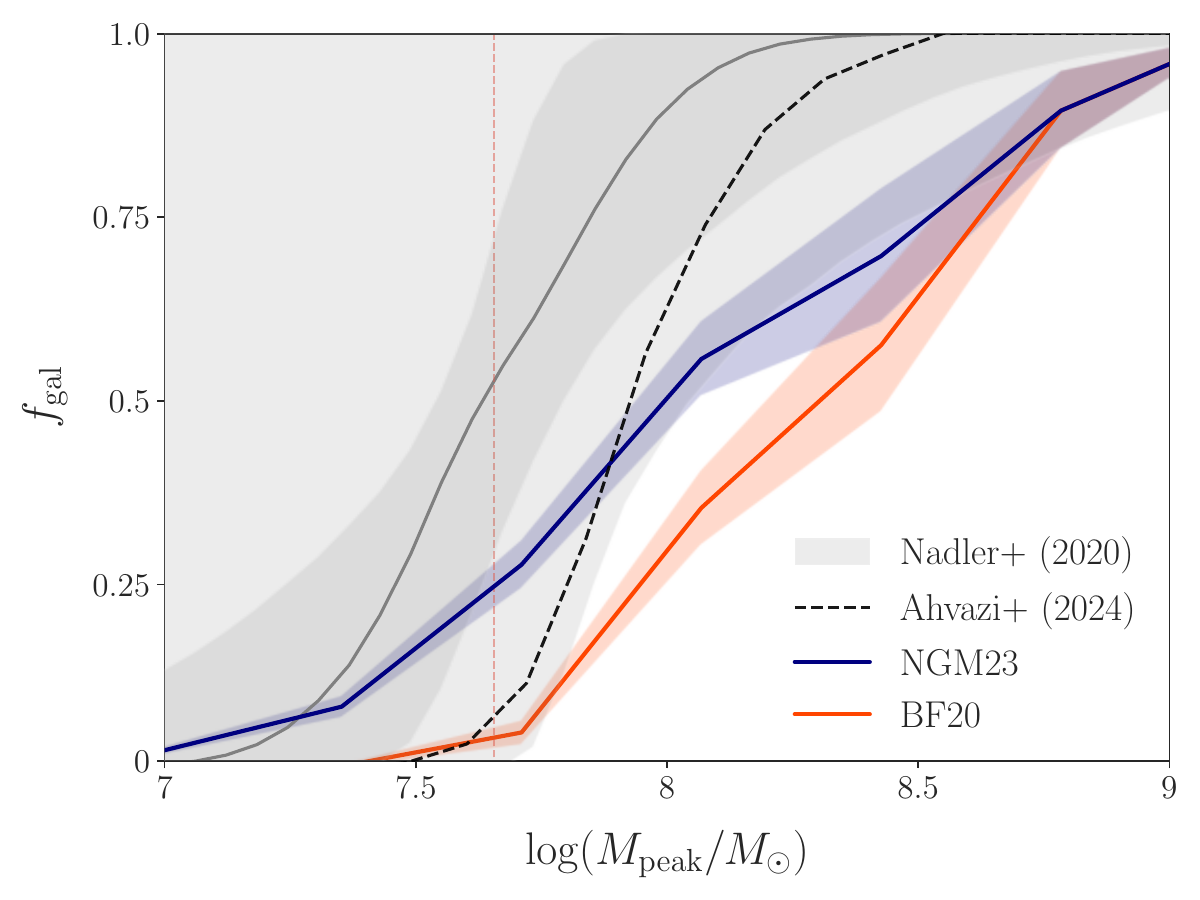}
\hspace{-1.25mm}
\includegraphics[trim={0cm 0.3cm 0 0},width=0.5\textwidth]{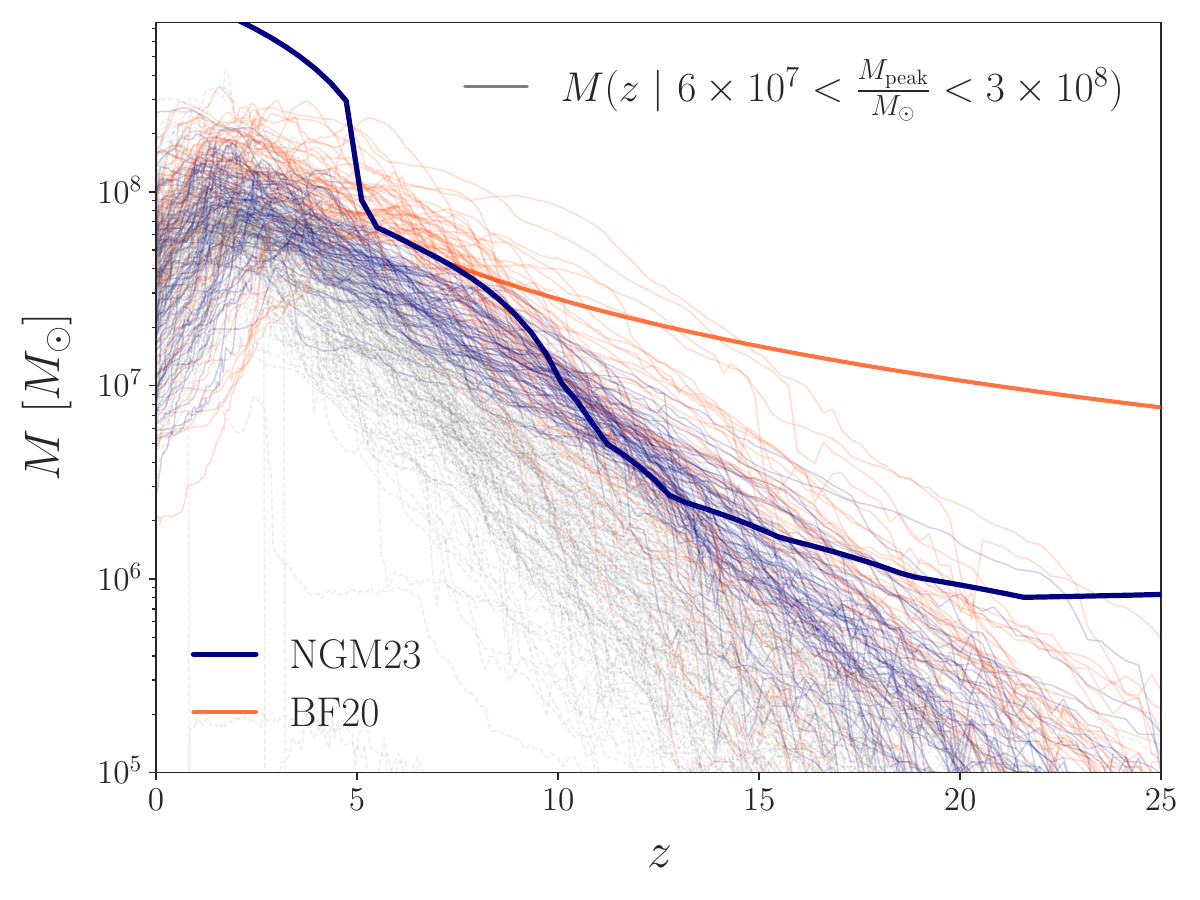}
\caption{\emph{Left}: Median galaxy occupation fraction for MW subhalos predicted using our MW analog simulation. Results are shown using the \citetalias{Nebrin230308024} (blue) and \citetalias{Benitez-Llambay200406124} (orange) $M_{\mathrm{crit}}(z)$ models. Shaded bands show $16\%$ to $84\%$ quantiles from bootstrap resampling. Dark (light) gray bands show the $68\%$ ($95\%$) confidence $f_{\mathrm{gal}}$ MW satellite posterior from \cite{Nadler191203302}; the thin gray line shows the median and the dotted vertical red line shows the resolution limit from that study. The dashed black line shows $f_{\mathrm{gal}}$ from the \cite{Ahvazi230813599} semianalytic model, which includes \HH\ cooling. \emph{Right}: MAHs for subhalos with $6\times 10^7~M_{\mathrm{\odot}}<M_{\mathrm{peak}}<3\times 10^8~M_{\mathrm{\odot}}$, colored according to whether they exceed $M_{\mathrm{crit}}$ from \citetalias{Benitez-Llambay200406124} (orange), \citetalias{Nebrin230308024} (blue), or neither (dashed gray), where both $M_{\mathrm{crit}}(z)$ models have been decreased by $\sigma_{\log(M)}$.}
\label{fig:fgal_simulation}
\end{figure*}

Including \HH\ cooling allows lower-mass halos to form stars and yields a shallower occupation fraction. To illustrate how this arises, the right panel of Figure~\ref{fig:fgal_Mcrit} shows isolated halo MAHs with $M_{\circ}=2\times 10^8~M_{\mathrm{\odot}}$ and $z_{\circ}=1$, where $f_{\mathrm{gal}}$ decreases rapidly when using the \citetalias{Benitez-Llambay200406124} $M_{\mathrm{crit}}(z)$ model and much more slowly in the \citetalias{Nebrin230308024} case. At this mass, the halos that exceed $M_{\mathrm{crit}}(z)$ from \citetalias{Benitez-Llambay200406124} are high-concentration objects that form early \citep{Wechsler0108151}. As $M_{\circ}$ decreases, MAHs decrease roughly linearly at all times according to Equation~\ref{eq:M_z}, but concentration only rises slowly ($c \sim M^{-0.1}$ or shallower; \citealt{Correa150200391}). Thus, when using a lower $M_{\mathrm{crit}}(z)$ threshold due to \HH\ cooling, we find a population of halos that exceed $M_{\mathrm{crit}}(z)$ from \citetalias{Nebrin230308024} but not from \citetalias{Benitez-Llambay200406124}.

\section{Predictions for Milky Way Subhalos}
\label{sec:sim}

To predict $f_{\mathrm{gal}}$ for MW subhalos in a cosmological setting, we run a DM--only zoom-in resimulation of a host from the Milky Way-est suite \citep{Buch240408043}. Specifically, we resimulate Halo004 with six \textsc{MUSIC} \citep{Hahn11036031} refinement regions, a high-resolution particle mass of $6.3\times 10^3~M_{\mathrm{\odot}}$, and a Plummer-equivalent gravitational softening of $40~\mathrm{pc}~h^{-1}$. This host forms in a large-scale underdensity, decreasing the expected strength of the LW background and thus $M_{\mathrm{crit}}(z)$ at its location \citep{Kulkarni201004169,Incatasciato230108242}. At $z=0$, the host's virial mass is $1.03\times 10^{12}~M_{\mathrm{\odot}}$ and it contains a $1.5\times 10^{11}~M_{\mathrm{\odot}}$ LMC analog, which accretes at $z=0.09$ (lookback time $1.3~\mathrm{Gyr}$) and reaches a pericentric distance of $59~\mathrm{kpc}$ today. We save a large number of output snapshots starting at $z=99$, and we generate halo catalogs and merger trees using \textsc{Rockstar} and \textsc{consistent-trees} \citep{Behroozi11104372,Behroozi11104370}.

From this simulation, we use the main-branch MAHs of all surviving subhalos with $M(z=0)>1.9\times 10^6~M_{\mathrm{\odot}}$, yielding $2930$ total subhalos.\footnote{We find that the present-day subhalo mass function is converged above this $300$-particle limit, consistent with \cite{Nadler220902675}.} We measure $f_{\mathrm{gal}}$ as a function of peak mass before infall into any host, $M_{\mathrm{peak}}\equiv \max(M(z>z_{\mathrm{infall}}))$. To assess statistical uncertainties, we resample subhalos in each $\log(M_{\mathrm{peak}}/M_{\mathrm{\odot}})$ bin with replacement. Rather than marginalizing over intrinsic scatter as in Section~\ref{sec:results}, we calculate uncertainties by bootstrap resampling the simulated MAHs; these statistical uncertainties dominate for current comparisons to observed MW satellites. To account for the underdense environment of our MW analog, we decrease $\log(M_{\mathrm{crit}}(z))$ by $\sigma_{\log(M)}$ in both $M_{\mathrm{crit}}(z)$ models, i.e., we set $p(\Delta)=\delta(\Delta+0.2)$. We comment on this choice below.

Our high-resolution resimulation is necessary to capture the $f_{\mathrm{gal}}$ cutoff, which is sensitive to the MAHs of halos with peak masses above $\approx 10^7~M_{\mathrm{\odot}}$ when \HH\ cooling is included. We resolve such subhalos with thousands of particles at first infall, and their progenitors are typically resolved with $\gtrsim 100$ particles (corresponding to masses of $\approx 10^6~M_{\mathrm{\odot}}$) up to $z\approx 20$. Our $M_{\mathrm{crit}}(z)$ models enable star formation in halo masses above $\approx 2\times 10^6~M_{\mathrm{\odot}}$ at these redshifts, implying that our resolution is sufficient. We confirm this argument by showing that our $f_{\mathrm{gal}}$ predictions converge in Appendix~\ref{sec:convergence}.

The left panel of Figure~\ref{fig:fgal_simulation} shows the resulting $f_{\mathrm{gal}}$ predictions for MW subhalos using the \citetalias{Nebrin230308024} (blue) and \citetalias{Benitez-Llambay200406124} (orange) $M_{\mathrm{crit}}(z)$ models. We predict that \HH\ cooling enables star formation in subhalos with $10^7~M_{\mathrm{\odot}}\lesssim M_{\mathrm{peak}}\lesssim 10^{8}~M_{\mathrm{\odot}}$; again, the lower limit is primarily set by DM--baryon streaming. These systems remain dark when only \H\ cooling and reionization are modeled, consistent with our isolated halo results in Figure~\ref{fig:fgal_Mcrit}. The \citetalias{Nebrin230308024} and \citetalias{Benitez-Llambay200406124} predictions converge at high $M_{\mathrm{peak}}$ because our $M_{\mathrm{crit}}(z)$ models match each other after reionization.

Our $f_{\mathrm{gal}}$ prediction for MW subhalos is similar to our isolated halo result with $z_{\circ}=1$ when using the \citetalias{Benitez-Llambay200406124} model, but the cutoff shifts toward higher masses by $\approx 1~\mathrm{dex}$ in the \citetalias{Nebrin230308024} case. This shift persists when $z_{\circ}$ in the isolated halo calculation is sampled from our subhalos' $z_{\mathrm{peak}}$ distribution. The difference results from a relatively low abundance of subhalos with $10^6~M_{\mathrm{\odot}}\lesssim M\lesssim 10^7~M_{\mathrm{\odot}}$ at $z\gtrsim 10$, as shown by the MAHs in the right panel of Figure~\ref{fig:fgal_simulation}. This is not a resolution artifact, since we resolve even lower-mass halos at these redshifts. We study host-to-host variance in $f_{\mathrm{gal}}$ using lower-resolution simulations in Appendix~\ref{sec:host-to-host}, finding that this shift may partly be a statistical fluctuation associated with our particular MW analog. The discrepancy may also indicate that the analytic MAH model does not accurately describe our subhalos' MAH distribution prior to infall, as suggested by comparing the right panels of Figures~\ref{fig:fgal_Mcrit} and \ref{fig:fgal_simulation}. Alternatively, $M_{\mathrm{crit}}(z)$ may fluctuate downward by more than $\sigma_{\log(M)}$ in the large-scale underdensity inhabited by our MW analog. For example, the $50\%$ occupation mass shifts from $\sim 10^8~M_{\mathrm{\odot}}$ to $10^{7.75}~M_{\mathrm{\odot}}$ ($10^{8.5}~M_{\mathrm{\odot}}$) when $\Delta=-0.4$ ($\Delta=0$) is used rather than $\Delta=-0.2$ in the \citetalias{Nebrin230308024} case. By fixing $\Delta=-0.2$, we have assumed that $M_{\mathrm{crit}}(z)$ decreases by an amount comparable to its intrinsic scatter in the underdense environment of our MW analog. It will be important to refine this assumption both theoretically, in order to self-consistently capture the correlation between subhalo MAHs in a given environment and the large-scale LW background, and observationally, by incorporating constraints on the photoionization history and LW background in the local Universe.

We compare our results to $f_{\mathrm{gal}}$ from the \cite{Ahvazi230813599} semianalytic model, which combines ePS merger trees with a galaxy formation model that includes \HH\ cooling but does not include DM--baryon streaming (see the dashed black line in the left panel of Figure~\ref{fig:fgal_simulation}). Our predicted $50\%$ occupation mass using the \citetalias{Nebrin230308024} model differs from \cite{Ahvazi230813599} by only $\approx 0.1~\mathrm{dex}$, although our $f_{\mathrm{gal}}$ prediction is shallower than theirs. This difference in slope may indicate a larger spread in our subhalos' MAHs, which is expected given that our subhalos form in a cosmological environment and are stripped before entering the host, causing their MAHs to deviate from semianalytic predictions based on ePS merger trees. Building pre-infall evolution into semianalytic models will clarify the impact of this effect.

Finally, we compare to the $f_{\mathrm{gal}}$ posterior from \cite{Nadler191203302}, which was inferred by forward-modeling the MW satellite population observed by the Dark Energy Survey and Pan-STARRS1 \citep{Drlica-Wagner191203302}. Our prediction that includes \HH\ cooling and DM--baryon streaming is consistent with \cite{Nadler191203302} at the $2\sigma$ level, while the case that only includes \H\ cooling and reionization is inconsistent with \cite{Nadler191203302} for $M_{\mathrm{peak}}\lesssim 10^{8.5}~M_{\mathrm{\odot}}$. These results imply that at least some known ultrafaint MW satellites formed through \HH\ cooling and hint that star formation is even more efficient than in \citetalias{Nebrin230308024}; we discuss potential explanations below.

\section{Discussion and Outlook}
\label{sec:discussion}

We have shown that \HH\ cooling decreases the peak mass of the smallest halos that can form stars from $\sim 10^8~M_{\mathrm{\odot}}$ to $\sim 10^7~M_{\mathrm{\odot}}$. This conclusion follows from two basic features of our model: (i) at early times, the halo mass threshold for star formation is $\sim 2\times 10^6~M_{\mathrm{\odot}}$ when \HH\ cooling and DM--baryon streaming are included, and (ii) typical minihalos then grow by $\sim 1~\mathrm{dex}$ in mass. Conversely, we predict that there is a lower limit of $\sim 10^7~M_{\mathrm{\odot}}$ on the peak halo mass of any galaxy, and that this limit is primarily set by DM--baryon streaming.

These conclusions are fairly robust to our choice of semianalytic $M_{\mathrm{crit}}(z)$ model, provided that \HH\ cooling and DM--baryon streaming are included. To illustrate this, we consider the \citetalias{Hegde230403358} model for $M_{\mathrm{crit}}(z)$, which shares many features with \citetalias{Nebrin230308024} but (i) includes X-ray feedback and (ii) does not include reionization.\footnote{In the absence of radiative feedback and DM--baryon streaming, the \HH\ cooling threshold itself differs between these studies due to their respective gas density profile assumptions. This difference does not affect our predictions because we include DM--baryon streaming and LW feedback.} Despite these differences, these authors' $M_{\mathrm{crit}}(z)$ predictions are in broad agreement. In detail, \citetalias{Hegde230403358} found that X-ray feedback lowers $M_{\mathrm{crit}}(z)$ relative to \citetalias{Nebrin230308024} by $\approx 20\%$ at $z\lesssim 15$ because it photoionizes the IGM and catalyzes \HH\ formation. Thus, including X-ray feedback may enable star formation in slightly lower-mass halos than we populate. Meanwhile, \citetalias{Hegde230403358} predicted a rise in $M_{\mathrm{crit}}(z)$ similar to \citetalias{Nebrin230308024} at $z\approx 5$ due to LW feedback.

Meanwhile, cosmological hydrodynamic simulations reveal effects beyond those captured in existing semianalytic models for $M_{\mathrm{crit}}(z)$. For example, \cite{Kulkarni201004169} and \cite{Schauer181112920,Schauer200005663} presented simulations including LW radiation and/or DM--baryon streaming. With both effects included, these studies measured $M_{\mathrm{crit}}(z)\approx 10^6~M_{\mathrm{\odot}}$ at $z\approx 20$, comparable to our \citetalias{Nebrin230308024} model. \cite{Kulkarni201004169} also showed that the effects of LW radiation and DM--baryon streaming on $M_{\mathrm{crit}}(z)$ are not independent. semianalytic models for $M_{\mathrm{crit}}(z)$ do not capture such correlations, which may impact the environmental dependence of $f_{\mathrm{gal}}$. Other hydrodynamic simulations predict efficient hydrogen deuteride (HD) cooling in halos below the \HH\ threshold (e.g., \citealt{McGreer08023918,Hirano150101630}); the \citetalias{Nebrin230308024} model does not include this effect. These predictions depend on HD chemistry and cooling assumptions \citep{Glover08031768}, and recent work suggests that the abundance of HD cooling halos is likely small ($\lesssim 10\%$; \citealt{Lenoble240116821}). Nonetheless, it will be important to study whether HD cooling impacts the galaxy formation threshold.

Recently, \cite{Hicks240720429} found that \HH\ cooling is pervasive in their hydrodynamic simulation, even for halos above the \H\ cooling threshold at $z\gtrsim 12$. In this work, \HH\ cooling is promoted by cold gas flowing into minihalos along filaments, analogous to cold-mode accretion in higher-mass halos \citep{Keres0407095}. Thus, the assumption of virialized gaseous halos---which underlies semianalytic models for $M_{\mathrm{crit}}(z)$---may break down in a cosmological setting, increasing the efficiency of \HH\ cooling and potentially alleviating the slight $f_{\mathrm{gal}}$ tension for MW subhalos that we identify.

Our conclusions are broadly consistent with semianalytic galaxy formation models. Specifically, our finding that \HH\ enables star formation in (sub)halos $M_{\mathrm{peak}}\lesssim 10^8~M_{\mathrm{\odot}}$ agrees with \cite{Ahvazi230813599}, who modeled \HH\ cooling in \textsc{Galacticus}. Predictions using \textsc{GRUMPY} in \cite{Kravtsov210609724} and \cite{Manwadkar211204511} also support this result, although these authors populated subhalos slightly below $M_{\mathrm{peak}}=10^7~M_{\mathrm{\odot}}$. In addition, \cite{Kravtsov210609724} found that the $50\%$ occupation mass decreases by $\sim 1~\mathrm{dex}$ when $z_{\mathrm{reion}}$ decreases from $9$ to $6$. Meanwhile, \cite{Chen220201220} used \textsc{A-SLOTH} to predict how $f_{\mathrm{gal}}$ depends on the DM--baryon streaming velocity; their results are consistent with our findings in Appendix~\ref{sec:streaming}. These authors found that other parameters in \textsc{A-SLOTH} (e.g., the Pop II initial mass function slope) do not affect $f_{\mathrm{gal}}$.

The occupation fraction is challenging to predict using cosmological hydrodynamic simulations because it probes galaxy formation over a wide range of spatial and temporal scales. Thus, it is not surprising that $f_{\mathrm{gal}}$ varies systematically with resolution in hydrodynamic simulations (e.g., \citealt{Munshi210105822}). For example, with a baryonic mass resolution of $30~M_{\mathrm{\odot}}$, \cite{Wheeler181202749} found that \emph{all} halos with $M(z=0)\gtrsim 4\times 10^8~M_{\mathrm{\odot}}$ form stars. Meanwhile, recent simulations from \cite{Gutcke220903366} approach single-star mass resolution and indicate that halos below $M_{\mathrm{crit}}(z)$ from \citetalias{Benitez-Llambay200406124} form stars, consistent with our results. We caution that---even at such high resolution---sub-grid models for gas evolution and star formation may affect $f_{\mathrm{gal}}$.

Looking forward, it is exciting to consider how this work informs searches for the remnants of the first stars (see \citealt{Bromm13055178} and \citealt{Klessen230312500} for reviews). In particular, our framework predicts the descendant properties of star-forming minihalos in a cosmological setting, including their distribution in the MW (also see \citealt{Jeon170207355,Rossi240612960,Hartwig241005393}). Combining these predictions with stellar evolution models offers a promising avenue to probe the galaxy formation threshold through near-field cosmology, which we look forward to pursuing.

\acknowledgments
We are grateful to Niusha Ahvazi for sharing data from \cite{Ahvazi230813599}, Stacy Kim for encouraging our resimulation, and Viraj Manwadkar for sharing data from \cite{Manwadkar211204511}. We thank the referee for constructive feedback, Niusha Ahvazi, Shaun Brown, Viraj Manwadkar, Olof Nebrin, Mike Norman, and Martin Rey for comments on the manuscript, and Andrew Benson, Dusan Kere{\v{s}}, Stacy Kim, and Risa Wechsler for helpful discussions.

The resimulation presented here was conducted through Carnegie's partnership in the Resnick High Performance Computing Center, a facility supported by Resnick Sustainability Institute at the California Institute of Technology. This
work used data from the Milky Way-est suite of simulations, hosted at \url{https://web.stanford.edu/group/gfc/gfcsims/}, which were supported by the Kavli Institute for Particle Astrophysics and Cosmology at Stanford University, SLAC National Accelerator Laboratory, and the US\ Department of Energy under contract number DE-AC02-76SF00515 to SLAC National Accelerator Laboratory.

\bibliographystyle{yahapj}
\bibliography{references}

\appendix

\section{Dark Matter--Baryon Streaming}
\label{sec:streaming}

\begin{figure*}[t!]
\includegraphics[width=\textwidth,trim={0 0.4cm 0 0}]{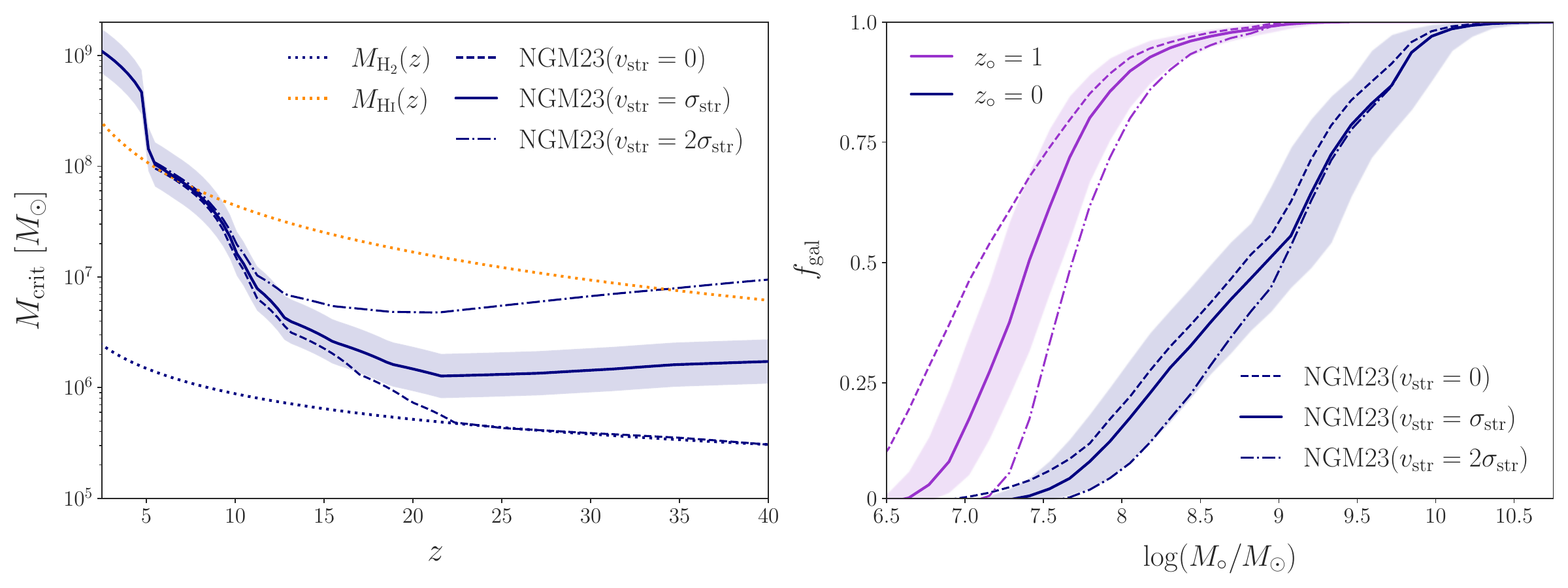}
\caption{\emph{Left}: Halo mass thresholds for star formation, showing versions of the \citetalias{Nebrin230308024} model with $v_{\mathrm{str}}=0$ (dashed), $v_{\mathrm{str}}=\sigma_{\mathrm{str}}$ (solid; our fiducial choice), and $v_{\mathrm{str}}=2\sigma_{\mathrm{str}}$ (dot-dashed). We only show $M_{\mathrm{crit}}(z)$ scatter for the $v_{\mathrm{str}}=\sigma_{\mathrm{str}}$ case for clarity, but we marginalize over this scatter in all cases. \emph{Right}: Median galaxy occupation fraction for isolated halos, predicted by combining analytic MAHs with the models in the left panel, with $z_{\circ}=1$ (purple) and $z_{\circ}=0$ (blue). Shaded bands show $16\%$ to $84\%$ intrinsic scatter quantiles for the $v_{\mathrm{str}}=\sigma_{\mathrm{str}}$ cases.}
\label{fig:fgal_stream}
\end{figure*}

Our fiducial results using the \citetalias{Nebrin230308024} model for $M_{\mathrm{crit}}(z)$ assume $v_{\mathrm{str}}(z)=\sigma_{\mathrm{str}}(z)$. The left panel of Figure~\ref{fig:fgal_stream} shows corresponding $M_{\mathrm{crit}}(z)$ models with $v_{\mathrm{str}}(z)=0$ (dashed) and $v_{\mathrm{str}}(z)=2\sigma_{\mathrm{str}}(z)$ (dot-dashed). As discussed in \citetalias{Nebrin230308024}, increasing $v_{\mathrm{str}}(z)$ raises $M_{\mathrm{crit}}(z)$, since it allows baryons to stream out of DM potentials and suppresses gas accretion rates. Decreasing $v_{\mathrm{str}}(z)$ has the opposite effect.

The right panel of Figure~\ref{fig:fgal_stream} shows our $f_{\mathrm{gal}}$ predictions for these models, generated using analytic MAHs. For reference, our fiducial results assuming $v_{\mathrm{str}}(z)=\sigma_{\mathrm{str}}(z)$ are shown by thick, solid lines for $z_{\circ}=1$ (purple) and $z_{\circ}=0$ (blue). The magnitude of $v_{\mathrm{str}}(z)$ clearly affects the $f_{\mathrm{gal}}$ predictions, such that lower-mass halos are occupied for smaller $v_{\mathrm{str}}(z)$. However, these shifts are fairly small compared to the difference between the $z_{\circ}=1$ and $z_{\circ}=0$ results, and the slope of $f_{\mathrm{gal}}$ is not significantly affected by variations in $v_{\mathrm{str}}(z)$.

\section{Mass--concentration Relation Scatter}
\label{sec:sigma_logc}

\begin{figure}[t!]
\hspace{-3.5mm}
\includegraphics[width=0.5\textwidth]{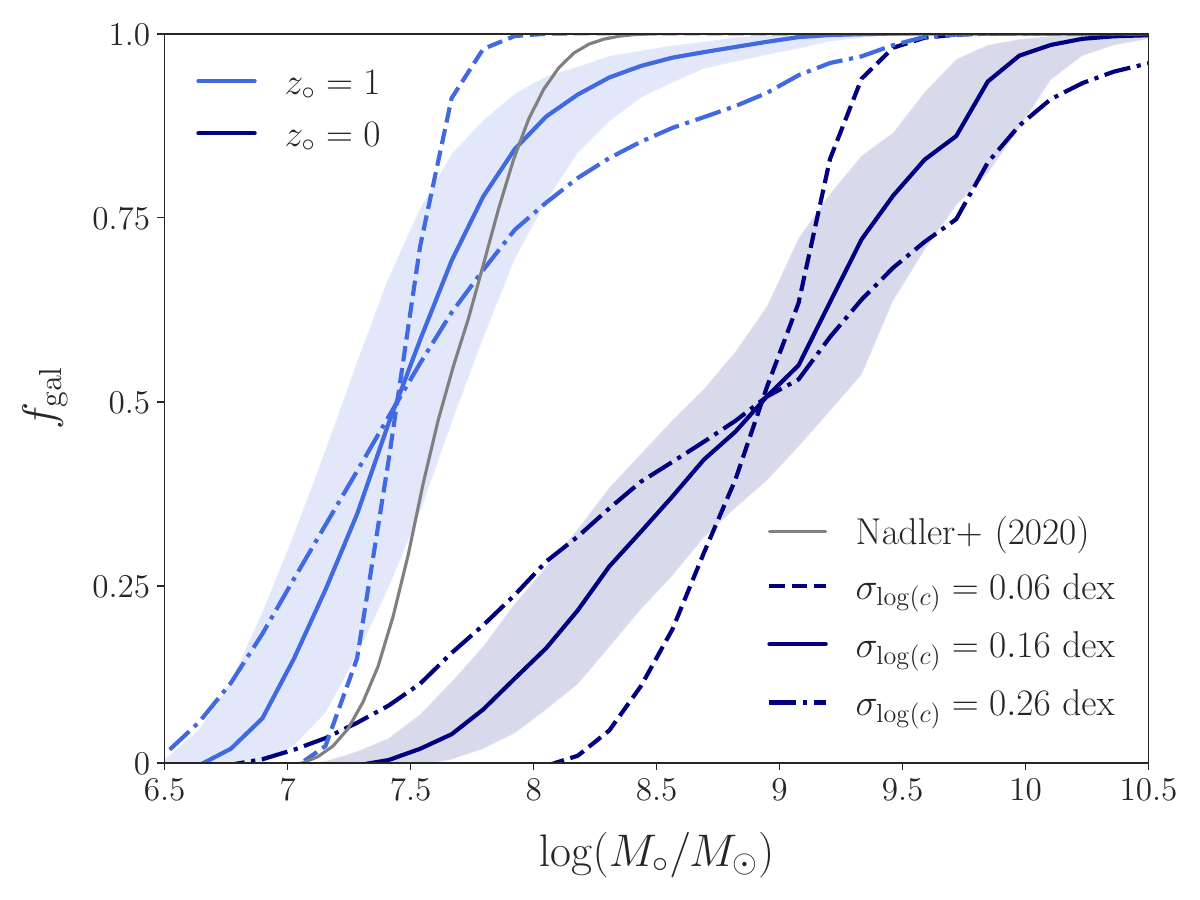}
\vspace{-6mm}
\caption{Median galaxy occupation fraction for isolated halos, predicted by combining analytic MAHs with our fiducial \citetalias{Nebrin230308024} $M_{\mathrm{crit}}(z)$ model, for $z_{\circ}=1$ (light blue) and $z_{\circ}=0$ (dark blue). Predictions are generated using $\sigma_{\log(c)}=0.06~\mathrm{dex}$ (dashed), $\sigma_{\log(c)}=0.16~\mathrm{dex}$ (solid; our fiducial choice), and $\sigma_{\log(c)}=0.26~\mathrm{dex}$ (dot-dashed). The thin gray line shows the best-fit model from \cite{Nadler191203302}, for reference. Shaded bands show $16\%$ to $84\%$ intrinsic scatter quantiles for the $\sigma_{\log(c)}=0.16~\mathrm{dex}$ cases; we marginalize over this scatter in all cases.}
\label{fig:fgal_sigma_logc}
\end{figure}

Our fiducial results assume $\sigma_{\log(c)}=0.16~\mathrm{dex}$, following the \cite{Diemer180907326} mass--concentration relation. Here, we consider how varying $\sigma_{\log(c)}$, as a proxy for scatter in the underlying distribution of halo MAHs, affects $f_{\mathrm{gal}}$ predictions. Figure~\ref{fig:fgal_sigma_logc} shows the results of this exercise using the analytic MAH model with $z_{\circ}=1$ (light blue) and $z_{\circ}=0$ (dark blue), for $\sigma_{\log(c)}=0.06~\mathrm{dex}$ (dashed), $\sigma_{\log(c)}=0.16~\mathrm{dex}$ (solid), and $\sigma_{\log(c)}=0.26~\mathrm{dex}$ (dot-dashed).

\begin{figure}[t!]
\hspace{-3.5mm}
\includegraphics[width=0.5\textwidth]{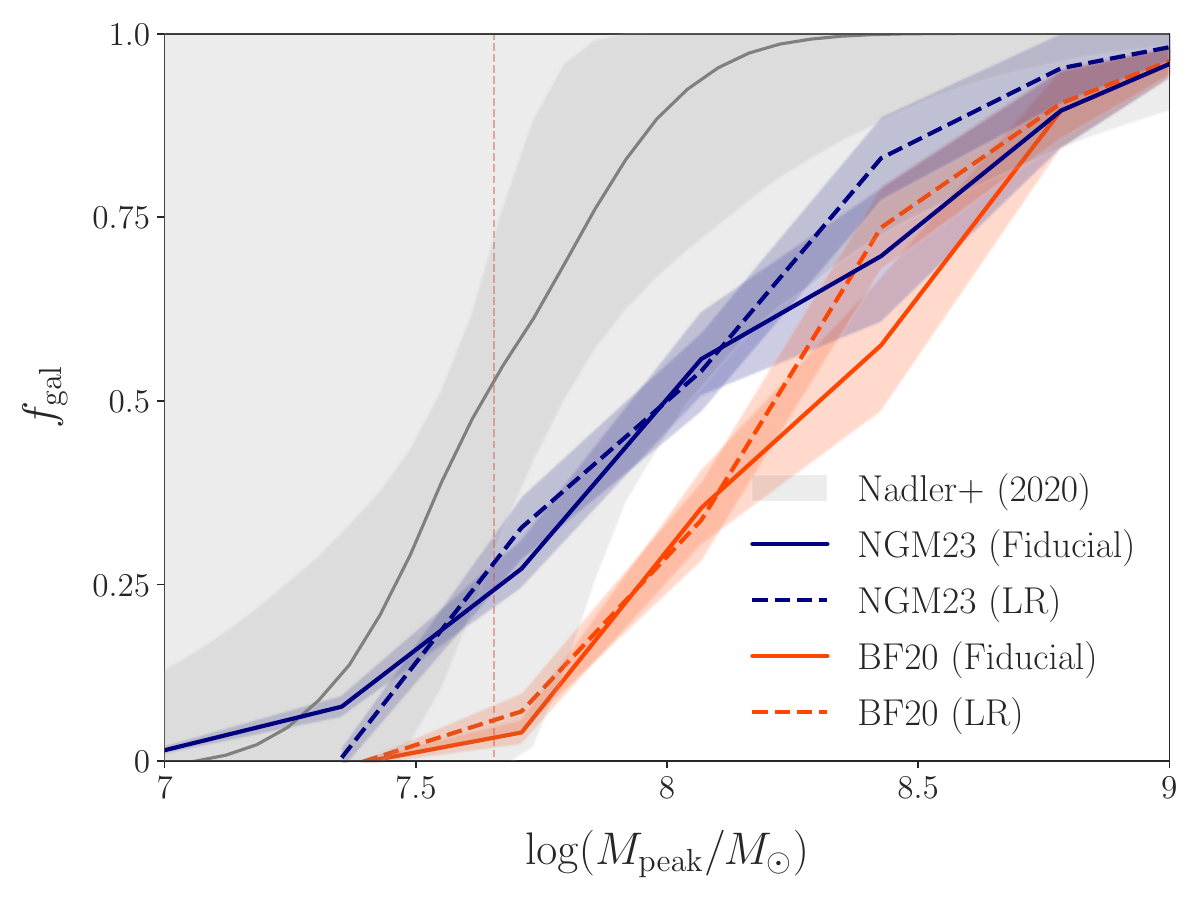}
\vspace{-6mm}
\caption{Comparison between our fiducial simulation predictions for $f_{\mathrm{gal}}$ ($m_{\mathrm{particle}}=6.3\times 10^3~M_{\mathrm{\odot}}$; solid) to a simulation of the same host at lower resolution (LR; $m_{\mathrm{particle}}=5\times 10^4~M_{\mathrm{\odot}}$; dashed). Results using the \citetalias{Nebrin230308024} model are shown in blue, and results using the \citetalias{Benitez-Llambay200406124} model with $z_{\mathrm{reion}}=7$ are shown in orange. Shaded bands show $16\%$ to $84\%$ quantiles from bootstrap resampling. The gray shaded region shows the \cite{Nadler191203302} posterior, as in Figure~\ref{fig:fgal_simulation}.}
\label{fig:fgal_simulation_convergence}
\end{figure}

The mass at which $f_{\mathrm{gal}}=50\%$ is nearly unchanged in all cases, and the $f_{\mathrm{gal}}$ cutoff becomes steeper as $\sigma_{\log(c)}$ decreases. This follows because, in the limit $\sigma_{\log(c)}\rightarrow 0$, MAHs become deterministic and $f_{\mathrm{gal}}$ approaches a step function at a mass determined by Equation~\ref{eq:M_z} and $M_{\mathrm{crit}}(z)$. We derive this dependence on $\sigma_{\log(c)}$ in Appendix~\ref{sec:derivation}.

\section{Occupation Fraction Derivation}
\label{sec:derivation}

Suppose that a halo with mass $M_{\circ}$ at redshift $z_{\circ}$ must exceed a critical mass $M_{*}$ at redshift $z_*>z_{\circ}$ to form stars. Under this assumption, Equation~\ref{eq:fgal} can be rewritten as
\begin{equation}
    f_{\mathrm{gal}}(M_{\circ},z_{\circ}) = p(M(z_*\lvert M_{\circ},z_{\circ})>M_*).\label{eq:fgal_inequality}
\end{equation}
For $M_{\circ}<M_*$, $f_{\mathrm{gal}}=0$ for MAHs that strictly increase with time, as in our isolated halo MAH model. For $M_{\circ}>M_*$, inserting Equation~\ref{eq:M_z} into Equation~\ref{eq:fgal_inequality} and rearranging yields
\begin{equation}
    f_{\mathrm{gal}}(M_{\circ},z_{\circ}) = p\left(\log(c_{\circ})>\log\left[\frac{Sc_1}{\ln(M_{\circ}/M_*)}\frac{z_*-z_{\circ}}{(1+z_{\circ})^2}\right]\right),\label{eq:fgal_temp}
\end{equation}
 Since $\log(c_{\circ})$ is normally distributed, the cumulative distribution in Equation~\ref{eq:fgal_temp} can be rewritten as
\begin{equation}
    f_{\mathrm{gal}}(M_{\circ},z_{\circ}) = \frac{1}{2}\left[1+\mathrm{erf}\left( \frac{\mu_{\log(c)}(z_{\circ}) - \log(c_*)}{\sqrt{2}\sigma_{\log(c)}}\right)\right],\label{eq:fgal_c}
\end{equation}
where $\mu_{\log(c)}$ is the mean of the \cite{Diemer180907326} mass--concentration relation and $c_*$ is the value of $c_{\circ}$ that saturates the inequality in Equation~\ref{eq:fgal_temp}.

This derivation shows that the mass-concentration relation scatter sets the slope of $f_{\mathrm{gal}}$, while its amplitude sets the cutoff scale. In particular, $f_{\mathrm{gal}}$ becomes steeper as $\sigma_{\log(c)}$ decreases and approaches a step function as $\sigma_{\log(c)}\rightarrow 0$, consistent with our results in Appendix~\ref{sec:sigma_logc}. Equation~\ref{eq:fgal_c} matches our fiducial results reasonably well when $z_*$ is chosen to be the lowest redshift that affects $f_{\mathrm{gal}}$ in a given model (e.g., for $z_*=10$ and $M_{*}=2\times 10^7$ and $5\times 10^7~M_{\mathrm{\odot}}$ for the \citetalias{Nebrin230308024} and \citetalias{Benitez-Llambay200406124} models, respectively). A more accurate analytic model would require integrating over redshift rather than assuming $f_{\mathrm{gal}}$ is determined at $z_*$.

Equation~\ref{eq:fgal_c} is similar to the form of $f_{\mathrm{gal}}$ used in previous work (e.g., \citealt{Graus180803654,Nadler191203302}),
\begin{equation}
    f_{\mathrm{gal}}(M_{\mathrm{peak}}) = \frac{1}{2}\left[1+\mathrm{erf}\left(\frac{\log(M_{\mathrm{peak}}/M_{\mathrm{\odot}})-\mathcal{M}_{50}}{\sqrt{2}\sigma_{\mathrm{gal}}}\right)\right],
\end{equation}
where $f_{\mathrm{gal}}(\mathcal{M}_{50})\equiv 0.5$. We leave a study of the correspondence between these expressions to future work.

\section{Convergence Test}
\label{sec:convergence}

Figure~\ref{fig:fgal_simulation_convergence} compares our fiducial $f_{\mathrm{gal}}$ prediction for MW subhalos to a lower-resolution (LR) simulation of the same host, run with a particle mass of $5\times 10^4~M_{\mathrm{\odot}}$ and softening length of $80~\mathrm{pc}~h^{-1}$, originally presented in \cite{Buch240408043}. For the LR simulation, we apply a $300$-particle cut at $z=0$ corresponding to $M_{\mathrm{vir}}>1.5\times 10^7~M_{\mathrm{\odot}}$. The LR prediction is consistent with our fiducial result within the bootstrap uncertainties for $M_{\mathrm{peak}}\gtrsim 1.5\times 10^7~M_{\mathrm{\odot}}$, while the high-resolution simulation resolves star formation in lower-$M_{\mathrm{peak}}$ subhalos when using the \citetalias{Nebrin230308024} model.\footnote{In the \citetalias{Nebrin230308024} case, the downturn in the LR $f_{\mathrm{gal}}$ result at high $M_{\mathrm{peak}}$ results from the slight increase in $M_{\mathrm{crit}}(z)$ from \citetalias{Nebrin230308024} relative to \citetalias{Benitez-Llambay200406124} near $z_{\mathrm{reion}}$. However, this downturn is not statistically significant.} We predict slightly higher $f_{\mathrm{gal}}$ in the LR simulation for $10^8\lesssim M_{\mathrm{peak}}/M_{\mathrm{\odot}}\lesssim 10^{8.5}$, which may result from noisy $M_{\mathrm{peak}}$ measurements at high redshifts in the LR simulation. Nonetheless, our $f_{\mathrm{gal}}$ predictions are not very sensitive to resolution down to the $300$-particle limit imposed throughout our analyses.

\section{Host-to-host Scatter}
\label{sec:host-to-host}

Here, we compare $f_{\mathrm{gal}}$ predictions for MW subhalos using several zoom-ins at a resolution lower than our fiducial resimulation. This comparison is meant to assess host-to-host scatter in $f_{\mathrm{gal}}$ and does not supersede our fiducial high-resolution result. Specifically, we compare $f_{\mathrm{gal}}$ predicted using the LR resimulation of Halo004 from Appendix~\ref{sec:convergence} to the Halo416 simulation at the same resolution from \cite{Nadler191203302}.\footnote{The main analyses in \cite{Nadler191203302} use an even lower-resolution version of Halo416, although they show that their main results converge when using the LR Halo416 run we consider here.} We also compare the LX14 Caterpillar-4 and Caterpillar-53 simulations analyzed in \cite{Manwadkar211204511}, which were run with a particle mass of $3\times 10^4~M_{\mathrm{\odot}}$ and a softening of $76~\mathrm{pc}~h^{-1}$, comparable to our Halo004 and Halo416 LR runs. Note that all four simulations we compare contain realistic LMC analogs. We apply a $300$-particle cut on $M(z=0)$ in each case, and we use the \citetalias{Benitez-Llambay200406124} model since the $f_{\mathrm{gal}}$ cutoff converges at low resolution in this case, according to Appendix~\ref{sec:convergence}. 

\begin{figure}[t!]
\hspace{-3.5mm}
\includegraphics[width=0.5\textwidth]{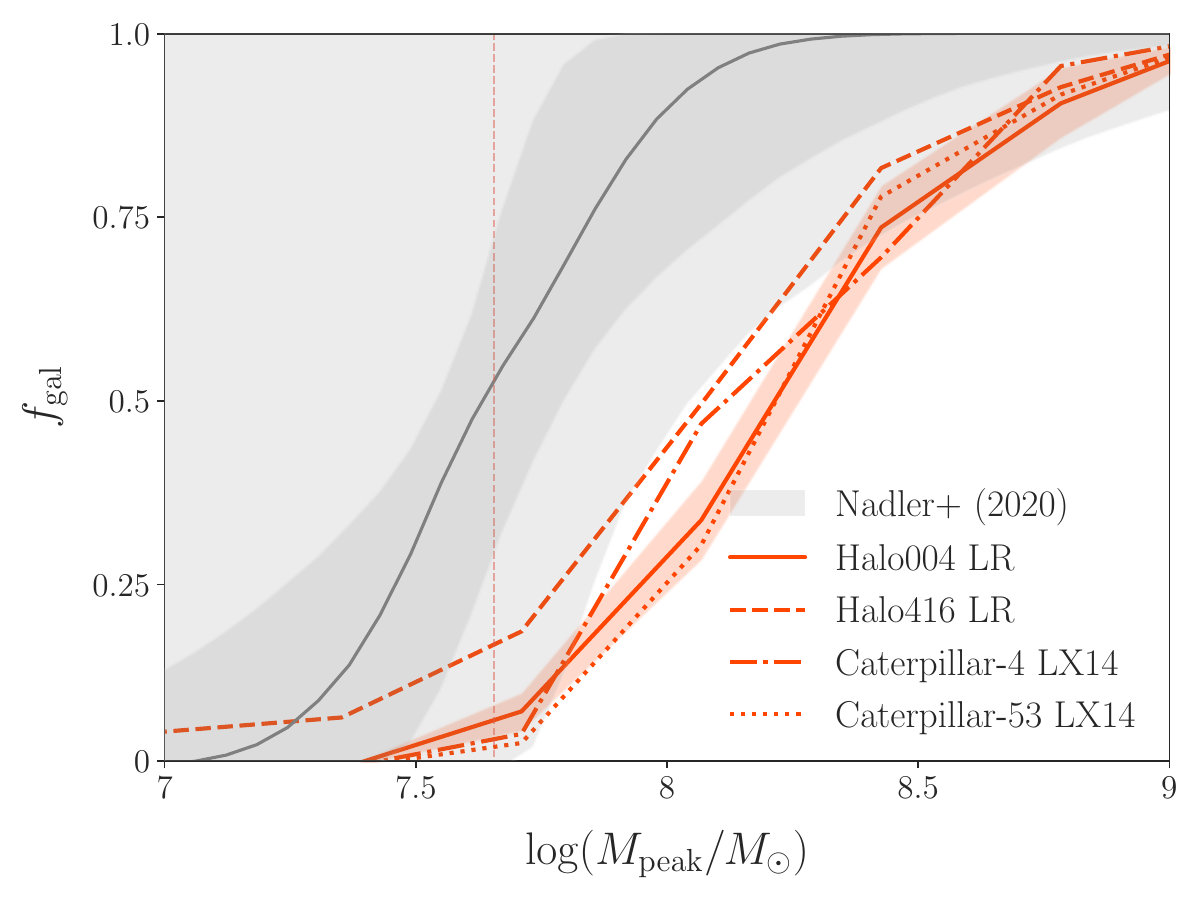}
\vspace{-6mm}
\caption{Comparison between $f_{\mathrm{gal}}$ predictions for the host used in this work (Halo004, solid), the host used in \cite{Nadler191203302} (Halo416, dashed), and the hosts used in \cite{Manwadkar211204511} (Caterpillar-4, dash-dotted, and Caterpillar-53, dotted). Halo004 and Halo416 are shown at the ``low resolution'' described in Appendix~\ref{sec:convergence}, which is comparable to the LX14 resolution of the Caterpillar hosts. Results are only shown for the \citetalias{Benitez-Llambay200406124} model (orange). The orange shaded bands shows $16\%$ to $84\%$ quantiles from bootstrap resampling the Halo004 LR result. The gray shaded region shows the \cite{Nadler191203302} posterior, as in Figure~\ref{fig:fgal_simulation}.}
\label{fig:fgal_simulation_comparison}
\end{figure}

Figure~\ref{fig:fgal_simulation_comparison} compares the resulting $f_{\mathrm{gal}}$ predictions. The Halo416 and Caterpillar occupation fractions are consistent with our Halo004 LR results when using the \citetalias{Benitez-Llambay200406124} model. In detail, slightly lower-mass halos are populated in Caterpillar-4 and Halo416 compared to Halo004. For Halo416, this shifts $f_{\mathrm{gal}}$ toward the \cite{Nadler191203302} posterior by an amount comparable to the bootstrap uncertainty on our Halo004 result. For all hosts except Halo416, virtually no halos with $M_{\mathrm{peak}}\approx 10^{7.5}~M_{\mathrm{\odot}}$ form stars in the \citetalias{Benitez-Llambay200406124} case. When using the \citetalias{Nebrin230308024} model, the host-to-host scatter in $f_{\mathrm{gal}}$ increases in this case, which is likely a result of poorly-resolved subhalo MAHs near the $f_{\mathrm{gal}}$ cutoff. We conclude that host-to-host scatter does not significantly affect our main conclusions, although it can ease the tension between our fiducial $f_{\mathrm{gal}}$ predictions and MW satellite constraints.

\end{document}